\documentstyle[pra,aps,twocolumn,epsfig]{revtex}

\begin{document}

\title{Quantum limit of optical magnetometry
in the presence of ac-Stark shifts}

\author{
 M.~Fleischhauer$^{1}$, A.B.~Matsko$^{2}$,
 and
 M.~O.~Scully$^{2,3}$
}
\address{%
 $^1$Sektion Physik,
 Universit\"at M\"unchen,
 D-80333 M\"unchen, Germany%
}
\address{%
 $^2$Department of Physics,
 Texas A\&M University,
 College Station, Texas 77843-4242,%
}
\address{%
 $^3$Max-Plank-Institut f\"ur Quantenoptik,
 Garching, D-85748, Germany,%
}

\date{\today}

\maketitle

 \tighten
\begin{abstract}
We analyze systematic (classical) and fundamental (quantum) limitations of
the sensitivity of optical magnetometers resulting from ac-Stark
shifts. 
We show that in contrast to absorption-based techniques, the signal 
reduction associated with classical broadening 
can be compensated in magnetometers based on phase measurements
using electromagnetically induced transparency (EIT). 
However due to ac-Stark associated  quantum noise the
signal-to-noise ratio of EIT-based magnetometers attains a maximum value 
at a certain laser intensity.
This value is independent on the quantum 
statistics of the light and defines a standard quantum limit of 
sensitivity. 
We demonstrate that an EIT-based optical magnetometer in Faraday 
configuration is the best candidate to achieve the highest sensitivity 
of magnetic field detection and give a detailed analysis of such a device.
\end{abstract}

\pacs{42.50.Lc, 07.55.-w,07.60.-j}

%%%%%%%%%%%%%%%%%%%%%%%%%%%%%%%%%%%%%%%%%%%%%%%%%%%%%%%%%%%%%%%%%%%%%

\section{Introduction}

%%%%%%%%%%%%%%%%%%%%%%%%%%%%%%%%%%%%%%%%%%%%%%%%%%%%%%%%%%%%%%%%%%%%%
\narrowtext

The detection of magnetic fields by optical means
is a well developed technique with applications ranging from
geology and medicine  \cite{Alexandrov}
to fundamental tests of violations of parity
and time-reversal symmetry \cite{parity}. 

In spite of their great variety,
optical magnetometers can be divided in two basic classes.
In the first class light absorption at a magnetic resonance 
is used to detect Zeeman level shifts, while the second class makes use of 
the associated changes of the index of refraction.
So called optical pumping magnetometer (OPM)
\cite{Alexandrov} as well as 
dark-state magnetometers 
based on absorption measurements \cite{nagel98} belong to
the first class.
The recently developed magnetometers
based on phase-coherent atomic media 
\cite{Scully92,Fleischhauer94} and the mean-field laser 
magnetometer of ref.\cite{Breitenaker92} belong to the second class.

If systematic measurement errors can be avoided, which
in practice can be a challenging task, 
the smallest detectable Zeeman shift (in units of frequency) 
is determined by the ratio of the noise level of the signal $S$
to its rate of change with respect to frequency
\begin{equation}
\Delta\omega_{\rm min} =\frac{S_{\rm noise}}
{\Bigl| {\rm d} S/{\rm d\omega}\Bigr| }.
\end{equation}
A fundamental lower limit of $S_{\rm noise}$ results from
photon counting errors due to shot-noise of the probe electromagnetic 
wave. 
$\bigl({\rm d}S/{\rm d}\omega\bigr)^{-1}$, which characterizes
a ``quality factor'' of the system,
is determined by an effective width of the magnetic resonance. 
The ultimate goal of magnetometer design is to minimize the
noise level and the effective width at the same time. 

The width of magnetic resonances in optical magnetometers is subject to 
two types of broadening: resonant power-broadening 
due to the coupling of the optical fields to the probe-transition
and a broadening due 
to ac-Stark shifts resulting from non-resonant couplings
to other transitions.
As shown in \cite{Scully92} and \cite{Fleischhauer94}  
power-broadening
 limits the
simultaneous minimization of noise and 
$\bigl({\rm d}S/{\rm d}\omega\bigr)^{-1}$
in {\it absorption} based magnetometers. In such devices 
increasing the probe laser power
reduces the shot-noise but does reduce the signal 
at the same time. As a consequence the sensitivity 
saturates at a rather low power level. On the other hand, as shown in 
\cite{Scully92} and \cite{Fleischhauer94}, this effect can be 
compensated in a magnetometer that detects {\it phase shifts}
of the probe electromagnetic wave propagating 
in an optically thick atomic medium under conditions of 
electromagnetically induced transparency (EIT) \cite{EIT}.
Theoretically a complete elimination is possible
in a 3-level $\Lambda$-type system.

In any real atomic system, however, there are non-resonant couplings
to additional levels which lead to ac-Stark shifts 
and an additional broadening of the magnetic
resonance proportional to the laser intensity.
In the present paper we analyze the influence of ac-Stark shifts
and show that they ({\it i}) can diminish
the magnetometer signal and ({\it ii}) lead to 
additional noise contributions. 
We show that in {\it absorption} based devices
ac-Stark broadening
leads to a further reduction of the signal.
In contrast it
only gives rise to a bias phase shift in
an phase-sensitive EIT magnetometer.
This bias phase shift 
can be calibrated but
is still a major source of systematic errors.
It can be eliminated, 
if an EIT magnetometer with Faraday configuration is considered.

 However,
in both, absorptive and dispersive type devices, ac-Stark shifts
give also rise to fundamental noise contributions which increase
with the laser power more rapidly than shot noise.
Hence the magnetometer sensitivity decreases
above a certain power level. The maximum value of sensitivity 
constitutes the standard quantum limit. 
For an EIT magnetometer based on phase-shift measurements 
this limit
is determined  by the dispersion-absorption ratio of the medium and the 
intensity-phase noise coupling due to the self-phase modulation
associated with ac-Stark shifts.

We also discuss the possibility
of further increasing the sensitivity by means of non-classical
light fields and show that the maximum sensitivity is
essentially independent of the light statistics.

The paper is organized as follows: In Sec.~II we discuss the
fundamental broadening mechanisms of magnetic resonances,
power-broadening and ac-Stark associated broadenings.
It is shown in Sec.~III that the classical signal reduction 
due to these broadenings can be compensated in phase-sensitive
EIT magnetometers
in contrast to absorption-based techniques.
In Sec.~IV  fundamental quantum noise sources are discussed
and the standard quantum limit of magnetometer sensitivity derived. 
A detailed analysis of an EIT-Faraday magnetometer is  given in Sec.~V
and the prospects of using non-classical input states are discussed.

%%%%%%%%%%%%%%%%%%%%%%%%%%%%%%%%%%%%%%%%%%%%%%%%%%%%%%%%%%%%%%%%%%%%%%%%
%%%%%%%%%%%%%%%%%%%%%%%%%%%%%%%%%%%%%%%%%%%%%%%%%%%%%%%%%%%%%%%%%%%%%%%%

\section{broadening of magnetic resonances}

%%%%%%%%%%%%%%%%%%%%%%%%%%%%%%%%%%%%%%%%%%%%%%%%%%%%%%%%%%%%%%%%%%%%%%%%
%%%%%%%%%%%%%%%%%%%%%%%%%%%%%%%%%%%%%%%%%%%%%%%%%%%%%%%%%%%%%%%%%%%%%%%%

Optical magnetometers measure in essence 
the position of certain resonances which are 
sensitive to magnetic level shifts. An important quantity that determines
the signal strength of such a measurement is the width of the
magnetic resonance. As a rule the narrower the resonance, the easier it is
to detect level shifts. 

Magnetic resonances with small natural width can be obtained
e.g. by coupling Zeeman or hyperfine  components of ground states
in atoms either with an RF field or via an optical  Raman transition. 
In an
optical magnetometer these ground-state sub-levels are then coupled
by laser fields to excited atomic states.
The optical coupling
is also used to detect energy shifts of the ground-state sub-levels
induced by a magnetic field. However, at the same time this coupling
leads to a broadening of the magnetic resonances 
via two  mechanisms:
({\it i}) {\it power-broadening} and ({\it ii}) {\it broadening due to
ac-Stark shifts}.

%%%%%%%%%%%%%%%%%%%%%%%%%%%%%%%%%%%%%%%%%%%%%%%%%%%%%%%%%%%%%%%%%%%%%%%%

\subsection{Power-broadening}

%%%%%%%%%%%%%%%%%%%%%%%%%%%%%%%%%%%%%%%%%%%%%%%%%%%%%%%%%%%%%%%%%%%%%%%%

The first mechanism is power-broadening due to the {\it resonant}
interaction  with the probe transition. 
When the Rabi-frequency $\Omega$ of the optical probe 
field exceeds the value
\begin{equation}
\Omega_{\rm crit}^{(1)}\sim\sqrt{\gamma\gamma_0},
\end{equation}
where $\gamma_0$ is the unbroadened
width of the magnetic resonance and
$\gamma$ the homogeneous linewidth of the optical transition,
the magnetic resonance becomes power-broadened.
(Here and below we assume that $\gamma \gg \gamma_0$.) 
The effective width scales linearly with the Rabi-frequency $\Omega$
of the optical field or the square root of the corresponding power
\begin{equation}
\Gamma_{\rm eff} = \gamma_0 + \alpha \, \sqrt{\frac{\gamma_0}{\gamma}}
\, |\Omega|+\cdots.
\end{equation}
$\alpha$ is some numerical pre-factor of order unity that depends on
the specific model \cite{Fleischhauer94,Fleischhauer95}.
This broadening effect leads to a substantial 
limitation of the signal in an optical pumping magnetometer,
as shown in \cite{Fleischhauer94} and \cite{Fleischhauer95}. 

%%%%%%%%%%%%%%%%%%%%%%%%%%%%%%%%%%%%%%%%%%%%%%%%%%%%%%%%%%%%%%%%%%%%%%%%

\subsection{Broadening due to ac-Stark shifts}

%%%%%%%%%%%%%%%%%%%%%%%%%%%%%%%%%%%%%%%%%%%%%%%%%%%%%%%%%%%%%%%%%%%%%%%%

The second broadening mechanism is due to {\it non-resonant} couplings
of the probe electromagnetic wave with other than the probe transition
and the associated ac-Stark shifts. 
The ac-Stark effect leads to a shift of the magnetic resonance
of 
\begin{equation}
\Delta\omega_{\rm ac-Stark} = \frac{|\Omega|^2}{\Delta_0}\label{omega_ac}
\end{equation}
where $\Delta_0$ is some effective detuning of non-resonant 
transitions from the frequency of the  probe field
weighted with relative oscillator strengths.
$\Omega$ is again the Rabi-frequency of the  probe field  
corresponding to the {\it resonant} probe transition.
($\Delta_0$ is of course just a model-dependent
coupling parameter.  We have used this notation here for simplicity 
of the discussions.)

 In the classical limit and for a homogeneous laser intensity
throughout the atomic vapor, there is only a constant frequency shift 
due to the ac-Stark effect. 
This shift can be calibrated.
However, maximum signal is usually achieved when the atomic
density is chosen such that there is a substantial absorption of the
probe field. Hence when the probe Rabi-frequency
exceeds the value
\begin{equation}
\Omega_{\rm crit}^{(2)} \sim\sqrt{\Delta_0\gamma_0}\label{crit2}
\end{equation}
the resonance frequency changes as a function of propagation
through the medium. This leads to an effective inhomogeneous
broadening of the magnetic resonance.
For example, the transmission of a cell
containing atoms with a Lorentzian magnetic resonance subject
to ac-Stark shifts is determined by the integrated 
imaginary part of the susceptibility ($\chi^{\prime\prime}=
{\rm Im}[\chi]$)
\begin{equation}
\int_0^L\!\! {\rm d}z\, \chi^{\prime\prime}(z)\sim
\int_0^L\!\! {\rm d}z\, \frac{\gamma_0}{\gamma_0^2+ (\Delta + 
|\Omega(z)|^2/\Delta_0)^2}.
\end{equation}
$|\Omega(z)|^2$ characterizes the $z$-dependent power of the probe field and
$\Delta$ the detuning from the un-shifted transition frequency.
It is easy to see, that there is a broadening of the magnetic resonance
depending on the magnitude
of the ac-Stark shifts and the details  of the absorption process. 
An important 
feature is that this broadening is proportional to the {\it square}
of the Rabi-frequency or the laser power. Thus above a certain
power level, determined by Eq.(\ref{crit2}) ac-Stark associated
broadening can exceed power broadening, which leads e.g. to further
reduction of the signal in an optical pumping magnetometer.

%%%%%%%%%%%%%%%%%%%%%%%%%%%%%%%%%%%%%%%%%%%%%%%%%%%%%%%%%%%%%%%%%%%%%
%%%%%%%%%%%%%%%%%%%%%%%%%%%%%%%%%%%%%%%%%%%%%%%%%%%%%%%%%%%%%%%%%%%%%

\section{compensation of  broadening effects 
         in EIT magnetometer}

%%%%%%%%%%%%%%%%%%%%%%%%%%%%%%%%%%%%%%%%%%%%%%%%%%%%%%%%%%%%%%%%%%%%%
%%%%%%%%%%%%%%%%%%%%%%%%%%%%%%%%%%%%%%%%%%%%%%%%%%%%%%%%%%%%%%%%%%%%%

We here demonstrate that the classical broadening mechanisms discussed
in the previous section do not necessarily lead to a reduction of the
magnetometer signal if phase measurement techniques are used.
It has been shown in detail in \cite{Fleischhauer94} 
and \cite{LukinPRL97}, that  power-broadening 
can be completely compensated in a phase measurement 
by making use of EIT 
in optically dense $\Lambda$-type systems.

The 3-level $\Lambda$ configuration of an EIT magnetometer
as well as the associated linear susceptibility spectrum of the
probe field are shown in Fig.1. 
Here and in the following we consider closed systems i.e. we
assume that there are no effective decay mechanism due to 
time-of-flight limitations.
The upper level of the probe-field
transition $|a\rangle \leftrightarrow |b\rangle$ is coupled to
a meta-stable lower level $|c\rangle$ by a coherent and
strong driving field of Rabi-frequency $\Omega_d$. 
The  probe field
Rabi-frequency is denoted as $\Omega_p$ ($\Omega_p\ll\Omega_d$)
and the coherence decay rate of the probe transition as $\gamma$.
$\Delta$ is the one-photon detuning of the drive field
and $\delta$ the two-photon detuning. 
The transverse decay rate 
of the two-photon resonance (magnetic resonance) 
is denoted as $\gamma_0$.
It is assumed that the corresponding population exchange between
the ground-state sub-levels is small and will be neglected

As in the case of
an OPM there is power-broadening 
in an EIT magnetometer as soon as 
$|\Omega_d|>\sqrt{\gamma\gamma_0}$.
A unique property of an EIT resonance is however that
the dispersion-absorption 
ratio of the {\it optical} transition
is given by the inverse of 
the width of the {\it ground-state} transition
$\gamma_0$ 
and is
{\it independent on the
drive power} if $|\Omega_d|>\sqrt{\gamma\gamma_0}$.
Under conditions of one-photon resonance ($\Delta=0$)  one finds 
for small two-photon detuning
\begin{eqnarray}
\chi^\prime \equiv {\rm Re}\, [\chi] &\sim & -\frac{\delta}
{|\Omega_d|^2+  \gamma \gamma_0},\\
\chi^{\prime\prime} \equiv {\rm Im}\, [\chi] &\sim & \frac{\gamma_0
}{|\Omega_d|^2+  \gamma \gamma_0}. 
\end{eqnarray}
The residual absorption at the EIT resonance
decreases with increasing laser power in the same way as the dispersion.
Thus in a phase shift measurement 
power broadening can be compensated by increasing the density
and keeping a constant optical depth of the medium.

Similarly one finds that as long as the drive-field Rabi-frequency
is large compared to probe-induced ac-Stark shifts, 
which is very well satisfied,
ac-Stark shifts of the magnetic resonance (eq.({\ref{omega_ac}))
lead only to a bias phase shift.
\begin{equation}
\Delta\phi_{ac-Stark} \sim \int_0^L\!\! {\rm d} z\, \frac{|\Omega(z)|^2 }{
\Delta_0},
\end{equation}
where $L$ is the length of the atomic vapor cell.
This phase shift can in principle be calibrated but 
gives rise to systematic
errors. As will be discussed in detail later on,
there is no such
bias phase shift in a resonant Faraday configuration.

We conclude this section by emphasizing that in phase-detection
schemes based on EIT the detrimental (classical) effects of 
power-broadening and
ac-Stark associated broadening are eliminated. 
In the following section we will discuss the fundamental quantum 
noise sources of such magnetometer schemes.

%%%%%%%%%%%%%%%%%%%%%%%%%%%%%%%%%%%%%%%%%%%%%%%%%%%%%%%%%%%%%%%%%%%%%
%%%%%%%%%%%%%%%%%%%%%%%%%%%%%%%%%%%%%%%%%%%%%%%%%%%%%%%%%%%%%%%%%%%%%

\section{quantum-noise limit of magnetic field detection via
 optical phase shifts
in the presence of ac-Stark effects}

%%%%%%%%%%%%%%%%%%%%%%%%%%%%%%%%%%%%%%%%%%%%%%%%%%%%%%%%%%%%%%%%%%%%%
%%%%%%%%%%%%%%%%%%%%%%%%%%%%%%%%%%%%%%%%%%%%%%%%%%%%%%%%%%%%%%%%%%%%%

The problem of sensitive detection of phase shifts is common in 
optics. On the quantum level, the sensitivity of such kind of 
measurements is restricted by ({\it i}) vacuum fluctuations in the system and 
({\it ii}) self-action noise due to nonlinearities in the system, as
for example caused by ac-Stark shifts.
The simultaneous presence of both noises usually leads
to an absolute limit of the sensitivity.

Let us discuss this  problem for 
the particular case of optical magnetometry
based on phase-shift measurements in an atomic medium.
The ultimate limit for the smallest detectable
phase shift  
is set by the generalized uncertainty relation \cite {Heisenberg}
between phase- $\Delta\phi\equiv \phi-\langle \phi\rangle $ \cite{remark} 
and photon-number fluctuations $\Delta n\equiv n-\langle n\rangle$
of the output field.
\begin{equation}
\langle \Delta \phi^2\rangle\langle \Delta n^2
\rangle \ge 1 + \frac{1}{4}\bigl\langle 
\{\Delta\phi, \Delta n\}\bigr\rangle^2,
\label{Heisenberg}
\end{equation}
where $\{,\}$ denotes the anti-commutator.
If phase- and photon-number fluctuations are uncorrelated, the second 
term on the r.h.s. vanishes and one recovers the familiar Heisenberg 
relation. 
In any real magnetometer schemes  phase and
intensity fluctuations are however correlated due to e.g.~ac-Stark 
shifts (self phase modulation), and thus the second term in 
Eq.(\ref{Heisenberg}) is in general 
nonzero. When the intensity-phase coupling is small, 
it can be characterized by
a linear coupling coefficient $\beta$ in the form 
$\Delta\phi = \Delta\phi_0+\beta\Delta n$, where $\Delta \phi_0$
denotes phase fluctuations not correlated to intensity fluctuations. 
Thus we find
\begin{equation}
 \langle\Delta\phi^2\rangle \ge
\frac{1}{\langle \Delta n^2
\rangle} +\beta^2 \langle \Delta n^2\rangle.
\end{equation}

The signal phase accumulated 
during the propagation through an
atomic vapor cell is proportional to the Zeeman splitting $\Delta\omega_B$,
the length of the cell $L$, and the 
dispersion of the real part of the susceptibility at the laser frequency
${\rm d}\chi^\prime/{\rm d}\omega$.
The cell length is restricted by the absorption at the laser frequency, and 
a reasonable upper limit for $L$ is the (amplitude) absorption length 
$L_{\rm abs}=(\pi\chi^{\prime\prime}/\lambda)^{-1}$.

 Thus the maximum
phase shift is
\begin{equation}
\Delta\phi\vert_{\rm max}
=\frac{1}{\chi^{\prime\prime}} \frac{{\rm d}\chi^\prime}{{\rm d}\omega}
\, \Delta\omega_B.
\end{equation}
One recognizes, that the sensitivity of phase measurements to 
Zeeman shifts is determined by the dispersion-absorption
ratio $\bigl(1/\chi^{\prime\prime}\bigr) \
{\rm d}\chi^\prime/ {\rm d}\omega$. 
% At a magnetic resonance this ratio
% is given by the inverse of its width $\Gamma_{\rm eff}$
%  which in general depends on the laser intensity,
% i.e. $\Gamma_{\rm eff}=\gamma_0 + \alpha
% \langle n\rangle +\cdots$.
 
The  limit for the smallest detectable Zeeman shift is
therefore given by
\begin{equation}
\Delta\omega_B\bigr\vert_{\rm min}= 
\left[\frac{1}{\chi^{\prime\prime}} 
\frac{{\rm d}\chi^\prime}{{\rm d}\omega}\right]^{-1} \Bigl[
\langle \Delta n^2
\rangle^{-1} +\beta^2 \langle \Delta n^2\rangle\Bigr]^{1/2}.
\end{equation}
Under the condition, that the dispersion-absorption ratio
is independent on the laser power,
% $\langle n\rangle \ll \gamma_0/\alpha $, i.e. for 
% a non-broadened magnetic resonance ($\Gamma_{\rm eff}\approx\gamma_0$)
the  r.h.s.~of this expression is minimized when
$\langle \Delta n^2\rangle\vert_{\rm opt} = \beta^{-1}$.
Therefore there is an absolute lower limit or ``quantum limit'' of magnetic 
field detection via phase-shift measurements
independent on the photon-number fluctuations
\begin{equation}
\Delta\omega_B\bigr\vert_{\rm min}= \left[\frac{1}{\chi^{\prime\prime}}
\frac{{\rm d}\chi^\prime}{{\rm d}\omega}\right]^{-1} \,\sqrt{2\beta}.
\label{limit}
\end{equation}
The absorption-dispersion ratio of a magnetic resonance is
usually given by its natural width, which can be rather small
if a two-photon Raman process between Zeeman- or hyperfine
components is used as in an EIT magnetometer. 

We will show later on that different measurement strategies
as well as the use of non-classical light fields do in general not
improve this result.

%%%%%%%%%%%%%%%%%%%%%%%%%%%%%%%%%%%%%%%%%%%%%%%%%%%%%%%%%%%%%%%%%%%%%%%
%%%%%%%%%%%%%%%%%%%%%%%%%%%%%%%%%%%%%%%%%%%%%%%%%%%%%%%%%%%%%%%%%%%%%%%

\section{EIT-based Faraday magnetometer}

%%%%%%%%%%%%%%%%%%%%%%%%%%%%%%%%%%%%%%%%%%%%%%%%%%%%%%%%%%%%%%%%%%%%%%
%%%%%%%%%%%%%%%%%%%%%%%%%%%%%%%%%%%%%%%%%%%%%%%%%%%%%%%%%%%%%%%%%%%%%%%

Let us now discuss in detail an EIT magnetometer in resonant nonlinear
Faraday configuration. 
For this we consider the propagation of a strong, linear polarized light
field through an optically dense medium, consisting of 
resonant $\Lambda$-type systems (atoms, quantum wells etc.) as shown in Fig.~2.
For simplicity we ignore optical pumping into lower states
other than those shown in the figure and assume a closed system.
For a resonant $J=1\to J=0$ transition (say), optical pumping into
the lower $m_J=0$ state depletes both states $m_J=\pm 1$ in the same
way and thus effectively diminishes the optical density but does not
affect the signal. Symmetric re-pumping can be used to maintain the
population in the relevant sub-system without affecting the
detection scheme. We include a dephasing of the ground-state coherence
with rate $\gamma_0$ and a population exchange rate between the
ground states $\gamma_{0r}$.

The two circular components $E_-$ and $E_+$ 
of the linear polarized light 
generate a coherent superposition (dark state) of the states 
$|b_\pm\rangle\equiv |J=1,m_J=\pm 1\rangle$.
A magnetic field parallel to the propagation axis leads to an 
anti-symmetric level shift of $|b_\pm\rangle$ and thus 
by virtue of the large linear dispersion at an 
 EIT-resonance to an opposite change
in the index of refraction for both components. As a result the
polarization direction is rotated, which is the so-called  
resonant nonlinear Faraday effect \cite{nl_Faraday}.
The difference to the  linear Faraday effect
is the presence of the intensity-dependent dark resonance generated 
by the action of the strong laser field as opposed to a usual absorption
resonance in the weak-field limit.
The rotation of the plane of polarization at the output can be measured by
detecting the intensity difference of two linear polarized
components $\pm$45$^{\rm o}$ rotated with respect to the
input polarization. 

An aspect of the system, which becomes
particularly important when strong fields are considered,
are non-resonant couplings of the two circular components
to other levels, which to lowest order give rise to ac-Stark shifts
of the states $|b_\pm\rangle$. In a Faraday configuration the ac-Stark 
shifts of $|b_+\rangle$ and $|b_-\rangle$ are exactly equal 
and opposite in sign due to symmetry and thus there is no
average effect on the signal and no bias
phase shift or rotation. 
Thus the Faraday magnetometer is not subject to systematic errors
associated with ac-Stark shifts.
However, as mentioned before, 
ac-Stark shifts cause a coupling
between intensity and phase fluctuations which need to be taken into
account.

%%%%%%%%%%%%%%%%%%%%%%%%%%%%%%%%%%%%%%%%%%%%%%%%%%%%%%%%%%%%%%%%%%%%%%%%%

\subsection{Detection scheme}

%%%%%%%%%%%%%%%%%%%%%%%%%%%%%%%%%%%%%%%%%%%%%%%%%%%%%%%%%%%%%%%%%%%%%%%%

We here consider the detection scheme shown in Fig.~3. 
A strong linear polarized
field initially polarized in $x$ direction 
propagates through a cell of length $L$ with
the magneto-optic medium. Due to the nonlinear Faraday effect the
plane of polarization is rotated by an angle $\phi/2$.

In order to detect this angle the intensity difference
of the two orthogonal output directions $1$ and $2$ is measured.
The operator for the number of counts is given by
\begin{equation}
{\hat n}= C \int_{t_m}\!\!{\rm d}t\, \Bigl( 
{\hat E}_2^{-}(t){\hat E}_2^{+}(t) - {\hat E}_1^{-}(t){\hat E}_1^{+}(t)
\Bigr).\label{hat_n}
\end{equation}
where ${\hat E}^{\pm}$ denote the positive and negative frequency part
of the output electric field operators, 
$t_m$ is the measurement time, and 
$C=2\epsilon_0  c A/\hbar \nu_0$, $A$ being the beam cross-section and
$\nu_0$ the resonance frequency. 
Making use of the field commutation relations
$[{\hat E}^{+}_{1,2}(L,t),{\hat E}^{-}_{1,2}(L,t^\prime)]=
C^{-1} \delta(t-t^\prime)$ and $[{\hat E}_1^{\pm},
{\hat E}_2^{\pm}]=0$, we can express the mean number of counts as well 
as the fluctuations in terms of normal-ordered correlation functions.
The latter allows to apply a c-number approach where the  operators
${\hat E}$ are approximated by stochastic complex functions
$E$. 
\begin{eqnarray}
\langle {\hat n}\rangle &=& \langle n_2\rangle-\langle n_1\rangle ,\\
\langle \Delta {\hat n}^2\rangle &=& \langle \Delta n^2\rangle 
+\langle n_1\rangle+\langle n_2\rangle.\label{Delta_n}
\end{eqnarray}
where $n_{1,2}$ follows form Eq.(\ref{hat_n}) by replacing the field operators
by c-numbers
\begin{equation}
n_{1,2} = C \int_{t_m} \!\!{\rm d}t\,  E_{1,2}^-(L,t) E_{1,2}^+(L,t).
\end{equation}

In the usual configuration only the $x$-polarized component
of the input field is excited and we will restrict the discussion
 to a vacuum input of the $y$-polarized component.
The propagation of the field through the
magneto-optical medium is most conveniently described in terms of right
and left circular components $E_\pm =(1/\sqrt{2})\bigl(E_x \pm i E_y\bigr)$, 
and we therefore have
\begin{equation}
{n}= -i C \int_{t_m}\!\!{\rm d}t\, \Bigl( 
{ E}_-^{-}{ E}_+^+ - { E}_+^{-}{ E}_-^+
\Bigr).\label{hat1_n}
\end{equation}
The propagation of the circular components can be characterized by
two parameters, the intensity transmission coefficient $\eta$
and the phase shift  $
\phi_\pm(L,t)$ of the respective component at the output.
\begin{eqnarray}
E_\pm^+(L,t) &=& E_\pm^+(0,t)\, \sqrt{\eta} \, {\rm e}^{i\phi_\pm(L,t)}.
\end{eqnarray}
In the limit of small magnetic fields the absorption of both
circular components is identical for symmetry reasons
i.e. there is no dichroism.
With this we obtain for 
cw-input fields
\begin{eqnarray}
\langle \hat n\rangle =\eta \langle n_x\rangle_{\rm in} 
\sin\phi_{\rm sig}
\approx \eta \langle n_x\rangle_{\rm in}\, \phi_{\rm sig},
\end{eqnarray}
where $\phi_{\rm sig}=\phi_+(L)-\phi_-(L)$ is the (stationary)  
signal phase shift. 
Similarly we can estimate the
fluctuations in lowest order of the small rotation angle $\phi$
in the case of an initially coherent field
\begin{eqnarray}
&&\langle \Delta{\hat n}^2\rangle = \eta\langle n_x\rangle_{\rm in}
+\eta^2\langle n_x\rangle_{\rm in}^2 \,\langle\delta\phi^2\rangle.
\label{noise}
\end{eqnarray}
The first term corresponds to the vacuum noise level and the second
term proportional to 
\begin{equation} \label{ass31}
\langle\delta\phi^2\rangle =
\frac{1}{t_m^2}\int\!\!\!\int {\rm d}t{\rm d}t^\prime\,
\langle\delta\phi(t),\delta\phi(t^\prime)\rangle
\end{equation}
describes fluctuations due to an intensity-phase noise coupling
in the medium. ($\langle a,b\rangle\equiv 
\langle[a-\langle a\rangle][b-\langle
b\rangle]\rangle$)

In the following  we calculate the loss factor $\eta$,
the signal phase shift $\phi_{\rm sig}$ and the fluctuations $\langle 
\delta\phi^2\rangle$ due to the 
intensity-phase noise coupling for
the EIT-Faraday magnetometer.

%%%%%%%%%%%%%%%%%%%%%%%%%%%%%%%%%%%%%%%%%%%%%%%%%%%%%%%%%%%%%%%%%%%%%%%%

\subsection{Medium susceptibility and field propagation}

%%%%%%%%%%%%%%%%%%%%%%%%%%%%%%%%%%%%%%%%%%%%%%%%%%%%%%%%%%%%%%%%%%%%%%%%

The stationary propagation of the right and left circular polarized 
electric field components through the atomic vapor
is described by  
Maxwell equations in slowly-varying amplitude and phase approximation
\begin{equation}
\frac{{\rm d}}{{\rm d}z} E_\pm^{+}(z) = \frac{i \nu_0}{2c \epsilon_0}
\wp_\pm N \sigma_{b\pm a}(z).
\end{equation}
$N$ is the atomic number density,
$\wp_\pm$ are the dipole moments of the respective transitions, 
 and $\sigma_{b\pm\,  a}$ are the c-number
analogues of the atomic lowering operators ${\hat \sigma}_{b\pm \, a} =
|b_\pm\rangle\langle a|$. Analytic expressions for 
$\sigma_{b\pm\,  a}$ can be obtained from the
stationary solution of the c-number Bloch equations for 
the atomic populations
\begin{eqnarray}
\dot \sigma_{b-\,  b-} &=& -\gamma_{0{\rm r}} (\sigma_{b-\,  b-}-\sigma_{b+\, b+}) 
+  \gamma_{{\rm r}} 
\sigma_{a\, a}\nonumber\\
&& -i(\Omega_-\sigma_{a\, b-}-c.c.), \label{sb-b-} \nonumber\\
\dot \sigma_{b+ \, b+} &=& \gamma_{0{ \rm r}} (\sigma_{b- \, b-}-\sigma_{b+ \, b+}) 
+  \gamma_{{\rm r}} 
\sigma_{a\, a} \nonumber\\
&& -i(\Omega_+\sigma_{a\, b+}-c.c.), \nonumber
\end{eqnarray}
and polarizations
\begin{eqnarray}
\dot \sigma_{a \, b\pm} &=& -\Gamma_{a \, b\pm} \sigma_{a \, b\pm} 
- i\Omega_{\pm}^*
(\sigma_{b\pm \, b\pm}-\sigma_{a\,a})\nonumber\\
&& -i \Omega_{\mp}^* \sigma_{b \mp \, b \pm}, \\
\dot \sigma_{b- \, b+} &=& -\Gamma_{b- \, b+} \sigma_{b- \, b+} - i \Omega_-
\sigma_{a \, b_+} + i \Omega_+^* \sigma_{b- \, a},\label{sab+-} 
\end{eqnarray}
where 
\begin{eqnarray}
\Gamma_{a \, b\pm} &\equiv& \gamma + {\gamma_{0{\rm r}} \over 2} + i \left (
\Delta +\delta_\pm \pm {\delta_0 \over 2} \right ), \\
 \Gamma_{b- \, b+} &\equiv&
\gamma_0 +\gamma_{0{\rm r}} + i \left(\delta_0+\delta_+-\delta_-\right).
\end{eqnarray}
$\gamma_{\rm r} $ is the  
radiative linewidth of the transitions $|a\rangle \to |b_\pm\rangle$,
and $\gamma $ is the homogeneous transverse
linewidth of the optical transitions 
$ | a \rangle \rightarrow | b_{\pm} \rangle  $.
$\delta_0$ is the Zeeman splitting and $\delta_\pm$ are the
ac-Stark shifts of levels $|b_\pm\rangle$. 
$\Omega_\pm$ are the complex Rabi-frequencies of the two optical
fields, $\Omega_\pm =\wp_\pm E^-_\pm/\hbar$.
We have disregarded Langevin noise forces in 
Eqs.(\ref{sb-b-}-\ref{sab+-}) associated with spontaneous emission
and collisional decay processes, since it was shown in 
\cite{Fleischhauer94} that atomic noises have a negligible
effect on the magnetometer sensitivity.

We calculate the stationary solutions of the Bloch-equations
by considering only the lowest order 
in $\gamma_0$, $\gamma_{0{\rm r}}$, $\delta_0$ and $\delta_\pm$.
In this limit we find
\begin{eqnarray}
\sigma_{ab_\pm} &=&\frac{ i \,\Omega_\pm 
(\gamma_0 |\Omega_\mp|^2+ \gamma_{0{\rm r}}|\Omega_\pm|^2)}
{|\Omega|^2(2\gamma (2 \gamma_{0{\rm r}}+\gamma_0 ) + 
|\Omega|^2)} \nonumber \\
&& -\Bigl(\delta_\pm\pm\frac{\delta_0}{2}\Bigr) \frac{2 \Omega_\pm 
|\Omega_\mp|^2 }{|\Omega|^2( 2 \gamma (2 \gamma_{0{\rm r}}+\gamma_0 )+ 
|\Omega|^2)} \label{sigma} \\
&&+ \frac{ \Delta}{\gamma} 
\frac{\Omega_\pm \left [\gamma_{0{\rm r}} (|\Omega_+|^4 + |\Omega_-|^4)
+2 \gamma_0 |\Omega_\mp|^2|\Omega|^2
\right ]}{|\Omega|^4 
( 2 \gamma (2 \gamma_{0{\rm r}}+\gamma_0 )+ |\Omega|^2)}
\nonumber
\end{eqnarray}
where $|\Omega(z)|^2=|\Omega_-(z)|^2+|\Omega_+(z)|^2$. 
Usually the coherence decay between the ground levels dominates the
population exchange and thus 
$ \gamma_0 \gg \gamma_{0{\rm r}}$. 

It is convenient to separately consider the spatial evolution of
amplitudes and phases
of the complex Rabi-frequencies
$\Omega_\pm(z) =|\Omega_\pm(z)| {\rm e}^{i\phi_\pm(z)}$.
The intensities of the two fields are attenuated in the same way
\begin{eqnarray}
\frac{{\rm d}}{{\rm d}z} |\Omega_\pm|^2
= -\kappa \frac{\gamma_0 \gamma_{\rm r}}{|\Omega|^2}
 \frac{|\Omega_+|^2|\Omega_-|^2}{
(2\gamma_0 \gamma + |\Omega|^2)
},\label{intensity}
\end{eqnarray}
where $\kappa=(3/4\pi) N\lambda^2$.

Eq. (\ref{intensity}) can be easily solved
when the  length $L$ of the cell is small enough, such that 
$|\Omega(L)|^2 \gg 2\gamma\gamma_0$. In the Faraday set-up discussed
here $\Omega_\pm(0)=\Omega(0)/\sqrt{2}$, and therefore 
$|\Omega_\pm(z)|^2 =|\Omega(z)|^2/2$.   
We thus arrive at
\begin{eqnarray}
|\Omega(z)|^2 &=&|\Omega(0)|^2
\Bigl(1-\frac{\gamma_0\gamma_{\rm r} \kappa z}{2|\Omega(0)|^2}
\Bigr)\nonumber\\
&=&|\Omega(0)|^2\Bigl(1-\alpha_0 z\Bigr).\label{intensity_sol}
\end{eqnarray}
It is interesting to note that under conditions of EIT the
residual  absorption is not exponential but linear. 
The intensity transmission coefficient is then given by 
\begin{equation}
\eta=\Bigl(1-\alpha_0 L\Bigr).
\end{equation}
The approximation  $|\Omega(L)|^2 \gg 2\gamma\gamma_0$ sets an upper limit 
for the losses, such that $1\ge\eta\gg 2\gamma\gamma_0/|\Omega(0)|^2$.

Similarly we find the phase equations
\begin{eqnarray}
\frac{{\rm d}}{{\rm d}z} \phi_- &=& 
\frac{\kappa \gamma_{\rm r} }{2 \gamma} 
\frac{\Delta \gamma_0 + \gamma (\delta_0/2-\delta_-)}
{2 \gamma_0 \gamma +   |\Omega|^2},
\label{phab1}
\\
\frac{{\rm d}}{{\rm d}z} \phi_+ &=& \frac{\kappa \gamma_{\rm r} }{2 \gamma} 
\frac{\Delta \gamma_0 - \gamma (\delta_0/2+\delta_+)}
{2 \gamma_0 \gamma +   |\Omega|^2}.
\end{eqnarray}
The contributions from the one-photon detuning $\Delta$
cancel when the relative phase  $\phi=\phi_+-\phi_-$ is considered
\begin{equation}
\frac{{\rm d}}{{\rm d}z} \phi =-{\kappa \gamma_{\rm r} \over 2} 
\Bigl(\frac{\delta_0}{|\Omega|^2} +\frac{\delta_+-\delta_-}
{|\Omega|^2}\Bigr).\label{phi}
\end{equation}
The first term describes the signal-phase shift due to a magnetic field
and the second term the ac-Stark contribution. 
Integration of Eq.(\ref{phi}) yields for the signal 
\begin{equation}
\phi_{\rm sig}=-\frac{\delta_0}{\gamma_0} 
{\rm ln}\,\left |   
{\Omega(0) \over \Omega (L)}
\right |^2 
\end{equation}
and the ac-Stark contribution
\begin{equation} \label{ass21}
\delta\phi(t)=-{\kappa \gamma_{\rm r} \over 2}
\int_0^L\!\!{\rm d}z\, \frac{\delta_+(z,t)-\delta_-(z,t)}{|\Omega(z)|^2}.
\end{equation}
%
%

%%%%%%%%%%%%%%%%%%%%%%%%%%%%%%%%%%%%%%%%%%%%%%%%%%%%%%%%%%%%%%%%%%%%%%%%%

\subsection{Ac-Stark shifts and associated noise}

%%%%%%%%%%%%%%%%%%%%%%%%%%%%%%%%%%%%%%%%%%%%%%%%%%%%%%%%%%%%%%%%%%%%%%%%%

Let us now discuss the average ac-Stark shift and the
corresponding noise contributions. For this we first consider
the effect of an off-resonant quantized field on the 
energy of a single atom in lowest non vanishing order
of perturbation. We then generalize the results for the average
ac-Stark shift and its fluctuations to an ensemble of atoms
by making the physically reasonable
assumption that ac-Stark shifts of different atoms are uncorrelated.

We decompose
the Hamiltonian of the single atom interacting with the 
quantized field in a rotating frame in the form
$H=H_0+H_{\rm S}$, where $H_0$ is the unperturbed part 
\begin{eqnarray}
H_0 &=& 
H_0^{\rm field}
+\hbar \omega_{b_-b_+} |b_-\rangle\langle b_-|
\nonumber\\
&&+\hbar \Delta_{ab_+}|a\rangle\langle a| +\hbar\sum_j
\Delta_j|c_j\rangle\langle c_j|.
\end{eqnarray}
$\Delta_{ab_+}=
\omega_{ab_+}-\nu_0$ and
$\Delta_j=\omega_{c_jb_+}-\nu_0$ are the detunings
 of the $|a\rangle -|b_+\rangle$ and $|c_j\rangle -|b_+\rangle$
transitions. 
\begin{eqnarray}
&&H_{\rm S} = -\wp_-| a \rangle \langle b_- |{\hat E}_-^+ 
  -\wp_+| a \rangle \langle b_+ |{\hat E}_+^+ \nonumber\\
&&\enspace
- \sum \limits_j 
\left(\wp_{j\, +} | c_j \rangle \langle b_+  |{\hat E}_+^+ +
\wp_{j\, -} | c_j \rangle \langle b_- | {\hat E}_-^+ \right)  
+ {\rm adj}.\label{hamv}
\end{eqnarray}
describes  the resonant and non-resonant couplings 
of the quantized fields to the atom.
The non-resonant couplings to the excited states $| c_j \rangle$
cause ac-Stark shifts. 
We here 
have assumed that both fields are nearly monochromatic
and have set the energy of level $|b_+\rangle$ equal to zero.
$\wp_{j \, \pm}$ are the dipole moments of the transitions $|c_j
\rangle  \rightarrow | b_{\pm} \rangle$.

We proceed by formally eliminating the excited states $|c_j \rangle$
 by means
of a canonical transformation in second order perturbation 
\begin{equation}
\tilde H = \exp (S) \, H \, \exp (-S)\simeq
H + \left [ S,H \right ] + \left [ S,  \left[ S, H \right ] \right ],
\end{equation}
where $S$ obeys the equation
\begin{equation}
\left [S, H_0\right ] = \sum \limits_j 
\left ( \wp_{j\, +} | c_j \rangle \langle b_+ | {\hat E}_+^+ +
\wp_{j\, -} | c_j \rangle \langle b_- | {\hat E}_-^+   + {\rm adj.} \right )
\end{equation}
Under conditions of exact
two-photon resonance for the fields 
 we obtain the transformation
operator 
\begin{equation} \label{s}
S=  \sum \limits_j \left ( {\wp_{j\, +} \over \Delta_j} 
|c_j \rangle  \langle b_+| {\hat E}_+^+ +
{\wp_{j\, -} \over \Delta_j} 
|c_j \rangle  \langle b_-| {\hat E}_-^+ - {\rm adj.} \right ) .
\end{equation}

Assuming that the population of all excited levels is  small, 
we eventually find for the transformed 
Hamiltonian 
\begin{eqnarray}
&&\tilde H \simeq H_0 -\wp_+   | a \rangle \langle b_+ |{\hat E}_+^+
-\wp_- | a \rangle \langle b_- |{\hat E}_-^+ 
\nonumber\\
&&- \sum \limits_j 
\left ( {\wp_{j\, +}^2 \over \hbar\Delta_j} 
|b_+ \rangle  \langle b_+| {\hat E}_+^-{\hat E}_+^+ +
{\wp_{j\, -}^2 \over \hbar \Delta_j} 
|b_- \rangle  \langle b_-| {\hat E}_-^-{\hat E}_-^+   \right ) 
\nonumber\\
&&- \sum \limits_j 
 {\wp_{j+}\wp_{j-} \over \hbar \Delta_j} \left (
|b_+ \rangle  \langle b_-| {\hat E}_+^-{\hat E}_-^+ + 
|b_- \rangle  \langle b_+| {\hat E}_-^-{\hat E}_+^+   \right ).
\end{eqnarray}

Let us assume now, that $\Delta_j$ is much larger than the natural 
width of the excited states, and therefore the  population transfer due to 
the non-resonant coupling  is negligible. 
We identify  $\sum_j
\wp_{j\,\pm}^2 /\Delta_j \to 
\wp^2 / \Delta_0$,
where $\Delta_0$ is some effective detuning. 
The dipole moments $\wp_{j+}$ and $\wp_{j-}$ 
have usually alternating signs for different
excited states $|c_\j\rangle$. We therefore set
$\sum_j\wp_{j+}\wp_{j_-}/\Delta_j=0$.
Then the ac-Stark shift of the single atom can be represented
by the operator expression
\begin{equation} \nonumber
\hat \delta_{\pm}^l(t) = {\wp^2 
\over \hbar^2 \Delta_0} \hat E_{\pm}^-(z_l,t)
\hat E_{\pm}^+(z_l,t),
\end{equation}
where $l$ specifies the atom and $z_l$  its location.
Thus we find for the average ac-Stark shift
\begin{equation}
\langle \hat \delta_{\pm}^l(t) \rangle = {\wp^2 \over \hbar^2 \Delta_0} 
\langle \hat E_{\pm}^-(z_l,t)
\hat E_{\pm}^+(z_l,t) \rangle =\frac{|\Omega(z_l,t)|^2}{2 \Delta_0},
\end{equation}
where $\wp^2|\langle \hat E_\pm(z_l,t)\rangle|^2/\hbar^2 =
\wp^2|\langle  E(z_l,t)\rangle|^2/2\hbar^2 = |\Omega(z_l,t)|^2/2$.
Similarly we obtain for the second-order moments of the
ac-Stark shifts $
\langle \hat x,\hat y\rangle
\equiv \langle 
\left(\hat x-\langle\hat x\rangle\right)
\left(\hat y-\langle\hat y\rangle\right) \rangle
$
\begin{eqnarray}
&&\langle \hat \delta_{\pm}^l(t) \hat \delta_{\pm}^l(t') \rangle
 \nonumber\\
&&\qquad={\wp^4 \over \hbar^4\Delta_0^2} 
\langle \hat E_{\pm}^-(z_l,t) \hat E_{\pm}^+(z_l,t)
\hat E_{\pm}^-(z_l,t') \hat E_{\pm}^+(z_l,t') \rangle,\\
&&\langle \hat \delta_+^l(t) \hat \delta_-^l(t') \rangle
 \nonumber\\
&&\qquad={\wp^4 \over \hbar^4\Delta_0^2} 
\langle \hat E_+^-(z_l,t) \hat E_+^+(z_l,t)
\hat E_-^-(z_l,t') \hat E_-^+(z_l,t') \rangle,
\end{eqnarray}
or after normal ordering
\begin{eqnarray}
&&\langle \hat \delta_{\pm}^l(t) \hat \delta_{\pm}^l(t') \rangle
= \nonumber\\
&&\enspace{\wp^4 \over\hbar^4 \Delta_0^2} \Bigl [
\langle  E_{\pm}^-(z_l,t) E_{\pm}^+(z_l,t)
 E_{\pm}^-(z_l,t') E_{\pm}^+(z_l,t') \rangle 
\nonumber\\
&&\qquad\quad+
{\delta(t-t') \over C} \langle  E_{\pm}^-(z_l,t) E_{\pm}^+(z_l,t) 
\rangle
\Bigr ], \label{fluct1}\\
&&\langle \hat \delta_+^l(t) \hat \delta_-^l(t') \rangle
= 
\nonumber\\
&&\enspace { \wp^4\over \hbar^4\Delta_0^2} \Bigl [
\langle  E_+^-(z_l,t) E_+^+(z_l,t)
 E_-^-(z_l,t') E_-^+(z_l,t') \rangle \Bigr]. \label{fluct2}
\end{eqnarray}
The first terms in Eqs.(\ref{fluct1}) and (\ref{fluct2}) correspond to 
classical fluctuations, while the second term in (\ref{fluct1}) 
is vacuum or shot noise. If the applied fields are in a coherent state
only the shot noise term survives. In any practical realizations
there are however large excess noise contributions and the first terms
are usually the dominant ones. We will show that
all  excess noise contributions are canceled in a Faraday magnetometer
and only the vacuum contribution  survives. 

We generalize the above single-atom results to an ensemble of atoms assuming
independent fluctuations of the ac-Stark shifts of different atoms,
i.e. 
\begin{equation}
\langle \hat \delta^j_\mu\hat\delta^k_\nu\rangle \sim \delta_{jk},
\end{equation}
where $\{\mu,\nu\}\in \{+,-\}$. 
We  introduce the continuous variable
\begin{equation} \label{cv1}
\hat \delta_{\pm}(z,t) = L \sum_j \delta(z-z_j) \hat \delta_{\pm}^j(t).
\end{equation}
In a continuum approximation, $\sum_j\to (1/L) \int_L\!{\rm d}z$, 
and we have
\begin{equation}
\langle \hat \delta_{\pm}(z,t) \rangle = {\wp^2 \over \hbar^2 \Delta_0} 
\langle \hat E_{\pm}^-(z,t)
\hat E_{\pm}^+(z,t) \rangle =\frac{|\Omega(z,t)|^2}{2 \Delta_0}.
\end{equation}
Similarly
\begin{eqnarray}
&&\langle \hat \delta_{\pm}(z,t), \hat \delta_{\pm}(z',t') \rangle =
 {L\wp^4\over \hbar^4 \Delta_0^2} 
 \delta (z-z') 
\nonumber\\
&&\enspace\times\Bigl [
\langle  E_{\pm}^-(z,t) E_{\pm}^+(z,t),
 E_{\pm}^-(z,t') E_{\pm}^+(z,t') \rangle
\nonumber\\
&&\qquad +
{\delta(t-t') \over C} \langle  E_{\pm}^-(z,t) E_{\pm}^+(z,t) \rangle
\Bigr ], \label{d2}
\end{eqnarray}
and
\begin{eqnarray}
&&\langle \hat \delta_+(z,t), \hat \delta_-(z',t') \rangle =
 {L\wp^4 \over \hbar^4 \Delta_0^2} 
 \delta (z-z') 
\nonumber\\
&&\enspace\times\Bigl [
\langle  E_{\pm}^-(z,t) E_{\pm}^+(z,t),
 E_{\pm}^-(z,t') E_{\pm}^+(z,t') \rangle\Bigr]. 
\end{eqnarray}
We here have used that 
in continuum approximation for 
any smooth function $f(z)$ holds 
\begin{equation}
L \sum_l \delta (z-z_l) \delta (z'-z_l) f(z_l) = \delta (z-z') f(z).
\end{equation}

It is now straight forward to
evaluate the quadratic deviation of the 
relative ac-Stark shift 
\begin{eqnarray} 
&&\left \langle \left (
{ \hat \delta_{+}(z,t) - \hat \delta_{-}(z,t) \over 
2 |\Omega (z)  |^2} \right ),\left (
{ \hat \delta_{+}(z',t') - \hat \delta_{-}(z',t') \over 
2| \Omega (z^\prime)  |^2} \right )
\right \rangle = 
\nonumber\\
&&\qquad \delta (z-z') \delta (t-t') 
{\wp^2 L \over 2\hbar^2 C \Delta_0^2  | \Omega (z)  |^2}.
\label{rss1}
\end{eqnarray}
We note that the classical excess noise contributions exactly cancel 
and only the vacuum contribution is left over. 
 Due to the intrinsic balancing in the
EIT-Faraday magnetometer excess noise contributions
are automatically canceled.
This is an important advantage of the Faraday  configuration 
as compared to the asymmetric EIT-magnetometer discussed in 
\cite{Scully92} and \cite{Fleischhauer94}.

Using Eqs.(\ref{ass31}), (\ref{ass21}) and (\ref{rss1})
we eventually find for the phase fluctuations 
due to ac-Stark shifts
\begin{equation}
\langle \delta \phi^2    \rangle = \frac 1{t_m}
\frac{\kappa^2 \gamma_{\rm r}^2 }{4 \Delta_0^2}
\frac{\wp^2 L}{\hbar^2 C} \int^L_0 \frac{1}{ |\Omega (z) |^2} dz
\end{equation}
%
%

%%%%%%%%%%%%%%%%%%%%%%%%%%%%%%%%%%%%%%%%%%%%%%%%%%%%%%%%%%%%%%%%%%%

\subsection{Signal-to-noise ratio and minimum detectable Zeeman shift}

%%%%%%%%%%%%%%%%%%%%%%%%%%%%%%%%%%%%%%%%%%%%%%%%%%%%%%%%%%%%%%%%%%%

The minimum detectable Zeeman shift is obtained by setting the 
mean number of counts 
\begin{equation}
\langle\hat n\rangle =  
\eta\, \langle n_x\rangle_{\rm in}\, \phi_{\rm sig}
=-\eta\, 
 \langle n_x\rangle_{\rm in}\,  \frac{\delta_0}{\gamma_0}
{\rm ln}\, \bigl[\eta^{-1}\bigr]
\end{equation}
equal to the quantum mechanical uncertainty
\begin{eqnarray}
&&\langle \Delta {\hat n}^2\rangle^{1/2} =
\Bigl[\eta\langle n_x\rangle_{\rm in} +\eta^2\langle n_x\rangle_{\rm in}^2 
\langle\delta\phi^2\rangle\Bigr]^{1/2} \nonumber\\
&&\enspace= \sqrt{\eta \langle n_x\rangle_{\rm in}}
\left[1 + \displaystyle \frac { |\Omega(0)|^4}{\Delta_0^2 \gamma_0^2} 
\eta \ (1-\eta) \ {\rm ln}  (\eta^{-1})
\right]^{1/2}
\end{eqnarray}
This yields the signal-to-noise ratio
\begin{equation} \label{delta_min} 
{\rm SNR} = \left[\frac{
\displaystyle \frac {\delta_0^2 }{\gamma_0^2} \langle n_x\rangle_{\rm in} 
\eta \
{\rm ln}^2 ( \eta^{-1} )}
{1 + \displaystyle \frac {|\Omega(0)|^4}{\Delta_0^2 \gamma_0^2} 
\eta \ (1-\eta) \ {\rm ln}  (\eta^{-1})}\right]^{1/2},
\end{equation}
which is maximized for an optimal power of the field
corresponding to
\begin{equation}
|\Omega(0)|^2_{\rm opt} = \sqrt {\frac{\Delta_0^2 \gamma_0^2 }
{ \eta \ (1- \eta) \  {\rm ln}  (\eta^{-1})}}\sim \Delta_0 \gamma_0.
\label{Omega_opt}
\end{equation}
Substituting the optimum Rabi-frequency (\ref{Omega_opt}) 
into (\ref{delta_min}) yields a maximum SNR for $\eta\approx 0.06$.
Thus we find the quantum limit for the detection of Zeeman level shifts
\begin{equation}
\delta_0^{\rm SQL} = \gamma_0\, f \left ( \frac{\gamma_r}{\Delta_0} 
\frac{3}{8 \pi}
\frac{\lambda^2}{A}
\frac{1}{\gamma_0 t_m} \right )^{1/2},
\end{equation}
where
\begin{equation}
f\equiv 
\left ( \frac{1-\eta}{\eta \  {\rm ln}^3  (\eta^{-1})} \right )^{1/4}
\end{equation}
 is a 
numerical factor which varies between 1 and 2 for $\eta=0.01
\cdots 0.8$. (Note that $\eta$ is the transmission coefficient under 
conditions of EIT. Without EIT the medium would be totally opaque.)
In Fig.~4 we have shown the minimum detectable Zeeman splitting
(proportional to the magnetic field) as function of the 
laser input power for different transmission coefficients. 

One clearly sees that for small laser powers shot-noise is dominant, while
for larger laser powers ac-Stark associated fluctuations take over. 
Also shown is the saturation behavior of an OPM
\cite{Fleischhauer94}. Due to power broadening
the sensitivity of an OPM saturates as soon as the Rabi-frequency
reaches the value $\sqrt{\gamma\gamma_0}$. 
In the EIT-Faraday magnetometer, on the other hand, the optimum Rabi-frequency
corresponding to the quantum limit is of the order of 
$\sqrt{\Delta_0\gamma_0}$.
Since $\Delta_0\gg\gamma$ much higher sensitivities can be achieved here.

%%%%%%%%%%%%%%%%%%%%%%%%%%%%%%%%%%%%%%%%%%%%%%%%%%%%%%%%%%%%%%%%%%%%%%%%%%%%

\subsection{Compensation of ac-Stark associated noise by use of 
non-classical input fields}

%%%%%%%%%%%%%%%%%%%%%%%%%%%%%%%%%%%%%%%%%%%%%%%%%%%%%%%%%%%%%%%%%%%%%%%%%%%%

It is well known, that the effect of self-phase modulation due to
refractive nonlinearities can be compensated, 
at least in part, by means of an optimum detection procedure 
(for example, by measuring not the phase, but
an appropriately chosen quadrature amplitude of the probe electromagnetic
wave)  and/or by using non-classical light 
\cite{Caves81,opt_measure}.
The properties of the input quantum state in the methods utilizing 
non-classical light are thereby chosen such 
that after the interaction the probe wave 
is in the coherent or phase-squeezed state. 

In the case of an optical magnetometer, 
ac-Stark shifts appear due to  non-resonant nonlinearities and it would 
seem that these shifts can in principle be compensated by an adapted 
measurement strategy and the use of non-classical light.
An essential condition for such methods is however 
that the system is nearly lossless
in order to preserve the non-classical state of light. 
On the other hand, as discussed above, 
the maximum signal in an optical magnetometer
is achieved under conditions of substantial absorption.
(We note that the SNR is proportional to ln$(\eta)^2$.)  
We will show in the following with simple estimates 
that this feature makes it impossible to
increase the sensitivity by using non-classical light.

Let us consider the simplest example of compensation
of ac-Stark associated noise by non-classical light.
We assume, that the slowly varying field operators in
the Heisenberg picture are  represented in the form
$\hat E_{\pm} = \langle \hat E \rangle
+\hat e_{\pm}$, where $\hat e_{\pm}$ is the fluctuation part.
To discuss the compensation of ac-Stark effects let us
disregard  the resonant coupling with the medium and the
associated absorption. Then  we find that the field fluctuations 
at the end of the vapor cell can be written as
\begin{eqnarray}
\hat e_-(t,L)= \hat e_-(t,0) +i\frac{\kappa \gamma_{\rm r} L}{2 \Delta_0} 
\Bigl[\hat e_-(t,0)+ \hat e_-^+(t,0)
\Bigr], \\
\hat e_+(t,L)=\hat e_+(t,0) -i\frac{\kappa \gamma_{\rm r} L}{2 \Delta_0} 
\Bigl[\hat e_+(t,0)+ \hat e_+^+(t,0)\Bigr].
\end{eqnarray}
The second terms in these equations are due to ac-Stark shifts. 
One can see
that the uncertainty of the phase difference increases
as a result of ac-Stark shifts, which 
leads to the sensitivity restriction, discussed above.

Let us assume now, that the incident field is
{\it squeezed} in
such a way, that the operators of the field fluctuations at
the input obey the relations
\begin{eqnarray}
\hat e_-(t,0)=\tilde e_- -i\frac{\kappa \gamma_{\rm r} L}{2 \Delta_0} 
(\tilde e_-+ \tilde e_-^+), \\
\hat e_+(t,0)=\tilde e_+ +i\frac{\kappa \gamma_{\rm r} L}{2 \Delta_0} 
(\tilde e_++ \tilde e_+^+).
\end{eqnarray}
Here $\tilde e_{\pm}$ are free-field operators (the corresponding 
state is the field vacuum), which
obey the commutation
relations $[{\tilde e}^{+}_{\pm}(t),{\tilde e}^{-}_{\pm}(t^\prime)]=
C^{-1} \delta(t-t^\prime)$ and 
$[{\tilde e}^{\pm}_{\pm}(t), {\tilde e}^{\pm}_{\pm}(t^\prime)]=0$.
Then, in the absence of losses, the effects of ac-Stark shifts
are completely compensated in the output and the 
output fields are coherent. 
\begin{eqnarray}
\hat e_-(t,L)=\tilde e_- , \\
\hat e_+(t,L)=\tilde e_+ .
\end{eqnarray}
The sensitivity of the phase measurement
would thus be determined by shot-noise only, 
$\Delta \phi = 1/ \sqrt {\langle n \rangle}$. 

In the absence of losses, the sensitivity of the detection can even be
better than the shot-noise limit, 
if the initial state of the field is appropriately chosen
\cite{opt_measure}.
Making use of a SU(2) Lie-group description, Yurke showed that
the sensitivity of a phase shift measurement in a
Mach-Zehnder interferometer  can approach the so-called 
Heisenberg limit $\Delta \phi \simeq 1/\langle n \rangle$, where 
$\langle n \rangle$ is the total 
number of registered quanta \cite{yurke86pra,kim98pra}.

However, in the presence of losses resulting from 
the resonant coupling the noise compensation 
by means of non-classical light is only
partial due to unwanted noises added by the medium.
Taking into account linear losses 
and assuming, that the 
entrance field is squeezed in the way discussed above,
we can rewrite the equation for the residual noises in the phase  as
follows:
\begin{equation}
\delta\phi(t)=-{\kappa \gamma_{\rm r} \over 2}
\int_0^L\!\!{\rm d}z\, \frac{\delta_+(z,t)-\delta_-(z,t)}
{|\langle \Omega (z)\rangle|^2} 
\sqrt{1-\eta(z)}.
\end{equation}
$\eta(z) = 1-\alpha_0 z$ is the $z$-dependent transmission coefficient.
The expression indicates, that for small losses in the medium, the noise
can be almost completely suppressed. A maximum signal is achieved however
when $\eta\ll 1$ and thus the use of non-classical light only leads to 
a marginal reduction of the ac-Stark associated noise.
This is in contrast to the measurement schemes discussed in 
\cite{Caves81,opt_measure} which utilize squeezing to improve sensitivity.
The change of the expression for the ac-Stark associated 
noise leads to a change of the sensitivity 
factor $f$ according to
\begin{equation}
f\longrightarrow \tilde f=
\left ( \frac{(1-\eta)({\rm ln} (\eta^{-1}) + \eta -1)}
{\eta \  {\rm ln}^4  (\eta^{-1})} \right )^{1/4}.
\end{equation}
It is easy to see, that 
$\tilde f \simeq f$ for all relevant values of $\eta$,
which means that squeezing does not improve the sensitivity of the 
detection.

The same conclusion
 can be drawn for any kind of optimal strategy of measurement to 
compensate ac Stark shifts. The main reason for this 
is that both, the magnitude of the signal and absorption losses 
increase with the density-length product of the atomic vapor cell.

%%%%%%%%%%%%%%%%%%%%%%%%%%%%%%%%%%%%%%%%%%%%%%%%%%%%%%%%%%%%%%%%%%%%%%%%%%%%

\section{Summary}

%%%%%%%%%%%%%%%%%%%%%%%%%%%%%%%%%%%%%%%%%%%%%%%%%%%%%%%%%%%%%%%%%%%%%%%%%%%%

We have 
discussed the influence of ac-Stark shifts on the sensitivity of
optical magnetometers. We have shown that these shifts cause
a broadening of the relevant  resonances and give rise 
to additional noise contributions.
In absorption-type magnetometers, such as OPMs, the ac-Stark
associated broadening as well as power-broadening lead to
a reduction of the signal. We have shown that the classical part of 
these effects 
can be completely compensated in an EIT magnetometer in Faraday 
configuration where polarization rotation 
or, equivalently, the relative phase shift of two circular components
is measured.

In a magnetometer based on phase measurements
ac-Stark shifts lead also to a coupling between intensity and phase
fluctuations. As a result there
are additional, ac-Stark associated fluctuations which 
dominate over shot noise 
beyond a critical laser power. For a certain optimum  intensity
the fundamental signal-to-noise ratio
attains a maximum value which
represents the standard quantum limit of optical magnetometer
based on phase-shift measurements.
This quantum limit is determined by the
dispersion-absorption ratio of the atomic medium and the
strength of the intensity-phase noise coupling. 
The unique property of EIT is to provide a dispersion-absorption ratio 
which is independent of power-broadening and 
is given by the lifetime  of a ground-state coherence. 
The minimum magnetic level shift corresponding to the quantum limit 
of EIT magnetometers can
thus be orders of magnitude smaller than that of  optical pumping
devices. 

We have shown that the best candidate to reach the standard
quantum limit is a magnetometer in Faraday configuration,
which has been analyzed in detail.
In an EIT-Faraday magnetometer the signal reduction due to 
power- and ac-Stark broadenings is compensated 
by large densities of the atomic vapor.
The influence of classical excess noise is completely
eliminated due to symmetry
 and there are much less sources for systematic errors. 
We have also shown that the use of non-classical light and different
detection techniques only marginally improves the attainable
sensitivity since a maximum signal is associated with substantial
losses in the atomic medium.

%%%%%%%%%%%%%%%%%%%%%%%%%%%%%%%%%%%%%%%%%%%%%%%%%%%%%%%%%%%%%%%%%%%%%%%%%%%%

\section*{Acknowledgements}

%%%%%%%%%%%%%%%%%%%%%%%%%%%%%%%%%%%%%%%%%%%%%%%%%%%%%%%%%%%%%%%%%%%%%%%%%%%%

The authors would like to thank M. Lukin for
stimulating  discussions
on the role of ac-Stark shifts. 
A.M. and M.O.S. gratefully acknowledge
further useful discussions with
Y.~Rostovtsev  and the support from
the Office of Naval Research, the National Science Foundation, 
the Welch Foundation,
the Texas Advanced Research and Technology Program and the Air Force
Research  Laboratories. 
M.F. gratefully acknowledges the financial support of the
Alexander-von-Humboldt foundation through the Feodor-Lynen
Program.

\newlength{\figwidth}
\setlength{\figwidth}{\textwidth} 
\multiply\figwidth by 4
\divide\figwidth by 10

%%%%%%%%%%%%%%%%%%%%%%%%%%%%  list of references %%%%%%%%%%%%%%%%%%%%%%%%%%%%%

%%%%%%%%%%%%%%%%%%%%%%  figures %%%%%%%%%%%%%%%%%%%%%%%%%%%%%%%%%%%%

%
%
\begin{figure}
\begin{center}
\leavevmode
\epsfxsize=6cm
\epsffile{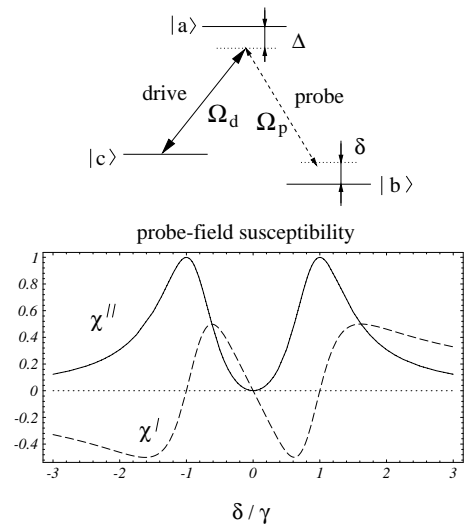}
\end{center}
\caption{Principle of a drive-probe EIT magnetometer.
Strong drive field in 3-level $\Lambda$ system (top) leads to 
transparency of probe field and linear dispersion around two-photon
resonance $\delta=0$ (bottom). 
Lower plot shows $\chi^\prime$ and $\chi^{\prime\prime}$ (real
and imaginary part of probe-field susceptibility) in arbitrary units
characterizing refractive index and absorption.
Drive-field Rabi-frequency equals natural width of probe transition.}
\end{figure}
\begin{figure}
\begin{center}
\leavevmode
\epsfxsize=6cm
\epsffile{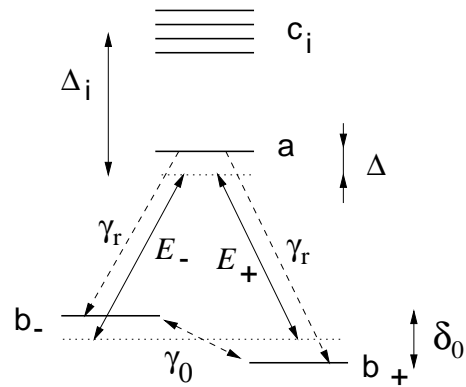}
\end{center}
\caption{$\Lambda$-system in Faraday configuration, 
$\gamma_r$ are radiative (longitudinal) decay rates, 
$\gamma_0$ the rate of ground-state 
coherence decay (transversal decay); 
$\Delta$ denotes one-photon- and $\delta_0$ 
magnetic-field induced two-photon detuning; $E_\pm$ describe
left- and right-circular polarized field
components. Population exchange (longitudinal decay) 
between ground-state sub-levels
is disregarded.
Also shown are non-resonant couplings to excited states
$|c_i\rangle$ causing ac-Stark shifts.}
\end{figure}
\begin{figure}
\begin{center}
\leavevmode
\epsfxsize=8cm
\epsffile{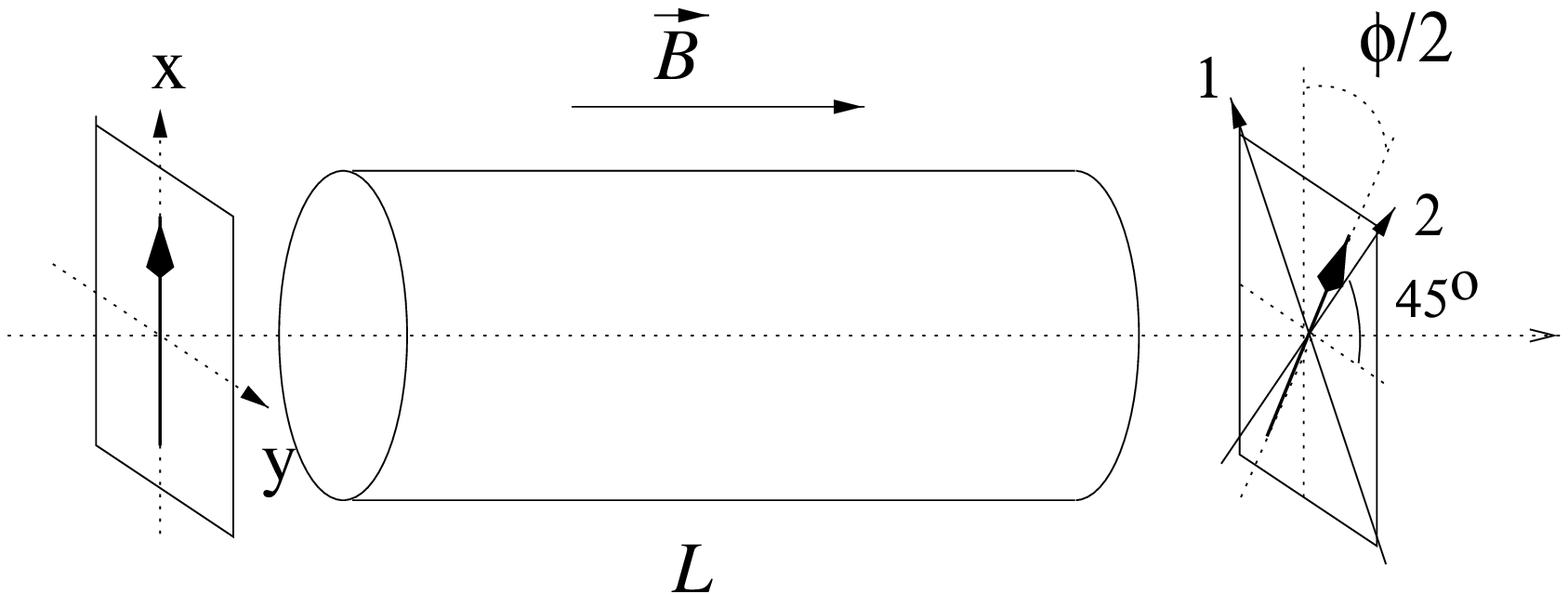}
\end{center}
\caption{Schematic drawing of Faraday measurement. 
Using polarizing beam-splitters the output field is
decomposed in two orthogonal components ${\hat E}_1$ and ${\hat E}_2$
$45^{\rm o}$ rotated relative to the $x-y$ system.}
\end{figure}
\begin{figure}
\begin{center}
\leavevmode
\epsfxsize=8cm
\epsffile{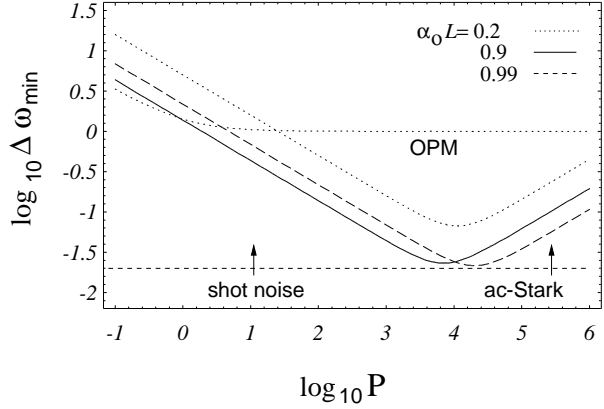}
\end{center}
\caption{Logarithm of minimum detectable Zeeman shift in arbitrary units 
as function of logarithm of laser power in units of 
$P_0= \hbar\nu_0 8\pi A/\lambda^2 \gamma_0$.
Transmission coefficients are $\eta=1-\alpha_0 L=0.8$, $0.1$, and $0.01$. $
\Delta_0/\gamma=10^3$. Also shown typical behavior of an optical pumping
magnetometer (OPM).}
\end{figure}
%
%

%%%%%%%%%%%%%%% figure captions %%%%%%%%%%%%%%%%%%%%%%%%%%%%%%%%%%%%%%%%%%%%%
%\newpage

%


\begin{references}
\bibitem{remark} Although there is no rigorous definition of a hermitian
phase operator, we nevertheless use it here noting that we restrict 
ourselves to states of the radiation field which have no significant overlap 
with the vacuum. In this case phase and photon number operators nearly 
coincide with the phase and amplitude quadrature operators.
\end{references}
\end{document}